\documentclass[12pt]{article}
\usepackage[dvips]{graphicx}

\newsavebox{\sboxpubnumber}
\newsavebox{\sboxpubdate}
\newcommand{\pubdate}[1]{\begin{lrbox}{\sboxpubdate}{#1}\end{lrbox}}
\newcommand{\pubnumber}[1]{\begin{lrbox}{\sboxpubnumber}{\begin{tabular}{l} #1 \\
				 \usebox{\sboxpubdate}
				 \end{tabular}}
                           \end{lrbox}
                           \pubblock}
\newcommand{\Title}[1]{\begin{center} {\Large #1 } \end{center}}
\newcommand{\Author}[1]{\begin{center}{ \sc #1} \end{center}}
\newcommand{\Address}[1]{\begin{center}{ \it #1} \end{center}}

\newcommand{\pubblock}{\rightline{
			\usebox{\sboxpubnumber}}}
\newenvironment{Abstract}{\begin{quotation}  }{\end{quotation}}
\newenvironment{Presented}{\begin{quotation} \begin{center}
             PRESENTED AT\end{center}\bigskip
      \begin{center}\begin{large}}{\end{large}\end{center}
      \end{quotation}}
\newcommand{\Acknowledgements}{\bigskip  \bigskip \begin{center} \begin{large}
             \bf ACKNOWLEDGEMENTS \end{large}\end{center}}

\newcommand{\BH}	{{\rm BH}}
\newcommand{\weak}	{{\rm W}}
\newcommand{\planck}	{{\rm pl}}
\newcommand{\CP}	{{\rm CP}}

\newcommand{\remnant}	{{\rm rem}}
\newcommand{\lnear}{{\begin{array}{c} < \\[-0.8em] \sim \end{array}}}

\newcommand{\etal}	{{\it et al.\ }}

\newcommand{\GeV}	{{\rm \;GeV}}

\newcommand{\keV}	{{\rm \;keV}}
\newcommand{\eV}	{{\rm \;eV}}
\newcommand{\kg}	{{\rm \;kg}}
\newcommand{\tons}	{{\rm \;tons}}

\newcommand{\second}	{{\rm \;second}}
\newcommand{\microgram}	{{\;\mu\rm g}}
\newcommand{\Gyear}	{{\rm \;Gyear}}

\begin{document}

\begin{titlepage}
\pubdate{December 7, 2001}	
\pubnumber{hep-ph/0112107}	

\vfill
\Title{%
Electroweak/GUT Domain Wall by Hawking Radiation:\\
Baryogenesis and Dark Matter\\
from Several Hundred kg Black Holes}
\vfill
\Author{Yukinori Nagatani}
\Address{Yukawa Institute for Theoretical Physics, Kyoto University\\
         Kyoto 606-8502, Japan}
\vfill
\begin{Abstract}
 A spherical domain wall around a small black hole is formed
 by the Hawking radiation from the black hole
 in the symmetry-broken-phase of the field theory, e.g.,
 the Standard Model (SM) and the Grand Unified Theory (GUT)
 which have a property of the phase transition.
 We have obtained two types of the spherical domain wall;
 (a) {\it thermalized wall}
 which is formed by the local heating up near black hole
 and symmetry restore locally
 and (b) {\it dynamical wall}
 which is formed by the balance
 between the pressure from the Hawking radiation
 and the pressure from the wall tensions.
 The electroweak wall is formed as a thermalized wall
 around a black hole with mass of the several hundred kilogram.
 The GUT wall is formed as a dynamical wall
 around much smaller black hole.
 The electroweak wall around a black hole can produce baryon number
 by the assumption of the CP-broken phase in the wall.
 The GUT wall can supply charge into the black hole,
 namely, the wall causes the {\it spontaneous charging up of the black hole}.
 We propose a cosmological model which can explain
 the origin of the baryon number and the cold dark matter
 by the primordial black hole with mass of the several hundred kilogram.
\end{Abstract}
\vfill
\begin{Presented}
    COSMO-01 \\
    Rovaniemi, Finland, \\
    August 29 -- September 4, 2001
\end{Presented}
\vfill
\end{titlepage}
\def\thefootnote{\fnsymbol{footnote}}
\setcounter{footnote}{0}

\section{Introduction}

The Hawking radiation
is a phenomenon of emitting particles
with a thermal energy-distribution characterized by the Hawking temperature
from a black hole \cite{Hawking:rv, Hawking:sw}.
In the case of a Schwarzschild black hole
which have neither charge nor spin,
the Hawking temperature is inversely proportional to its mass
and the total luminosity of the radiation
is inversely proportional to the square of the mass.
Then a small black hole injects
the intense Hawking radiation with high energy into the small region.
Actually the radius of the black hole with mass of several Hundred
kilogram is only $10^{-8}$ of the proton-radius, however,
the Hawking temperature is over $10^7\GeV$ and
the total luminosity is greater than the solar one.
Therefore the neighborhood of the small black hole gives us
the interesting experimental field of the particle physics
\cite{Cline:1996mk, Heckler:1995qq}.

\section{Thermalized Domain Wall by Black Hole}

\begin{figure}[htbp]%
\begin{center}%
 \begin{minipage}[b]{70mm}
  \includegraphics[scale=0.6]{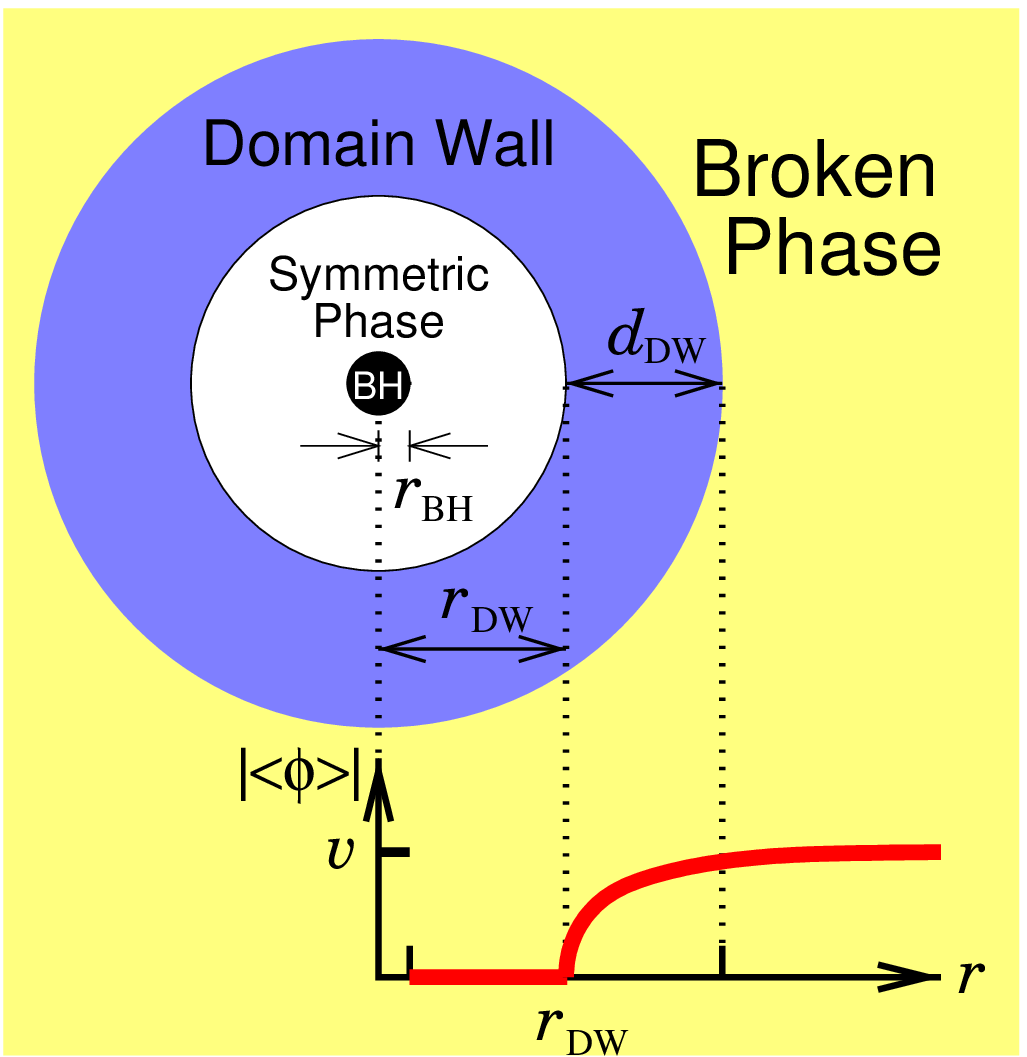}%
  \caption{Thermalized domain wall around a black hole and Higgs VEV.}
  \label{BHWall.eps}
 \end{minipage}
 \rule{10mm}{0mm}
 \begin{minipage}[b]{70mm}
  \includegraphics[scale=0.6]{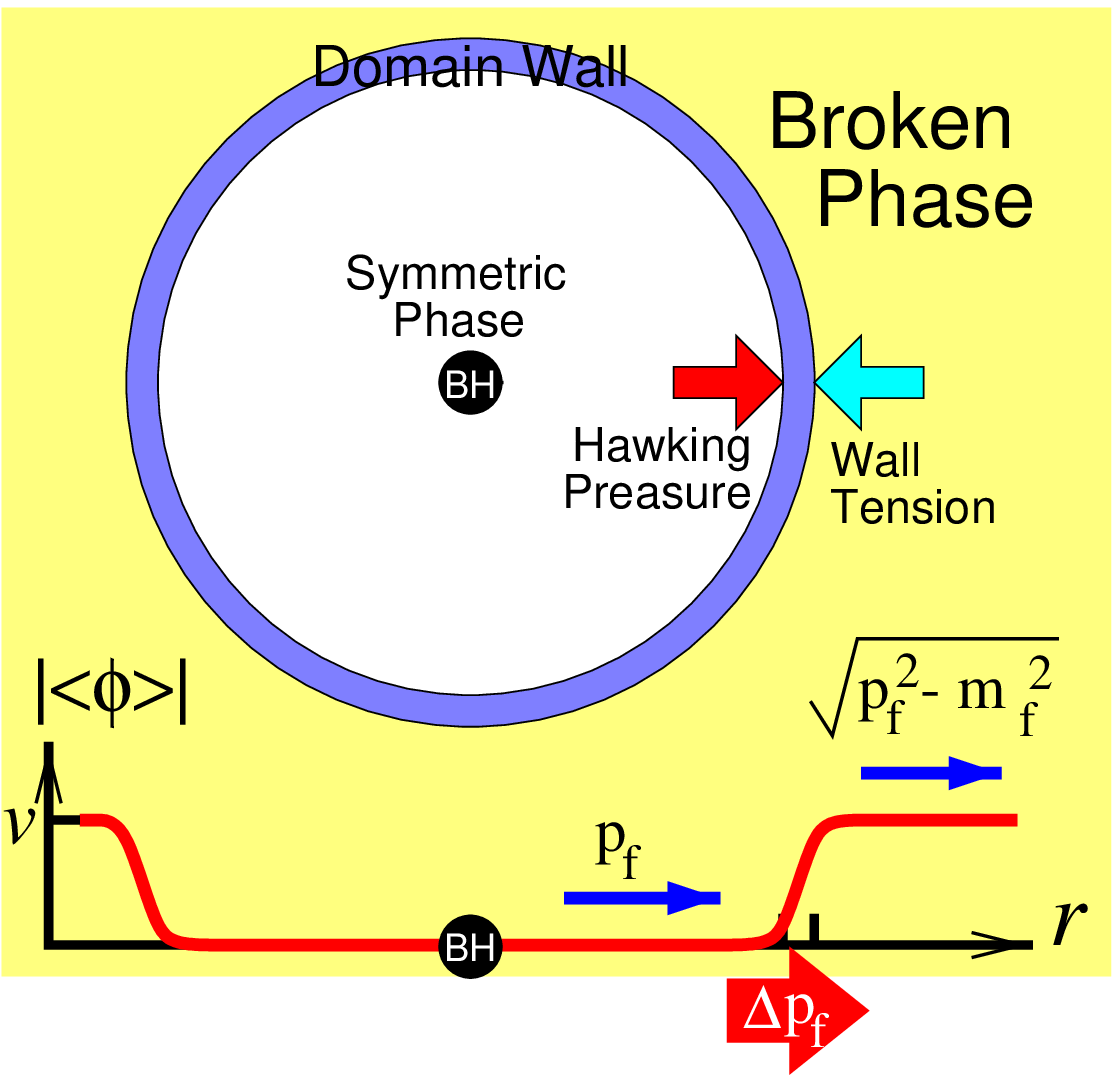}%
  \caption{Dynamical domain wall.}
  \label{NonThermal.eps}
 \end{minipage}
\end{center}%
\end{figure}

It is natural to ask what phenomenon is arisen
when we put a small black hole into
a broken phase vacuum of the field theory
which has a property of the (spontaneous) symmetry breaking,
e.g., the Standard Model (SM) and the Grand Unified Theory (GUT).
We can easily imagine that
the black hole heat up its neighborhood locally by the Hawking radiation
and the symmetry in the region is restored.
So, we can expect a formation of a spherical domain wall
whose center is the black hole and
which separates the symmetric phase region and the broken phase region
\cite{Nagatani:1998rt}.
First, we have analyzed the system in which
a black hole is radiating
at the electroweak-symmetry-broken-phase of the SM
by considering the {\it thermalization} with the interactions
and the {energy transfer} of all particles in the SM.
We have found that the electroweak (EW) domain wall
which is stationary and thermalized by the SM interactions
is formed when the mass of the black hole is greater than $70\kg$
and is smaller than $400\kg$ (or $200\tons$)
\cite{Nagatani:1998gv, Nagatani:2001nz}.
The black hole with larger mass than the upper limit
can not thermalize the wall by the SM interactions,
and the black hole with smaller mass than the lower limit
has been evaporated before the formation of the stationary wall.
One of the properties of our EW domain wall is that
the formation of the wall does not depend on
the order of the EW phase transition
because the locally heating up by the Hawking radiation
forms the wall.
The structure of the EW domain wall is
given by the thermalization near the black hole,
namely, it is given by 
\begin{eqnarray}
 \left<\phi(r)\right> &=& \left<\phi\right>_{T\;=\;T(r)},
\end{eqnarray}
where $T(r)$ is the temperature distribution around the black hole
and which is determined by the energy transfer analysis
with the local thermal equilibrium approximation.
We will show the structure of the EW domain wall in Figure \ref{BHWall.eps}.
We have assumed the EW phase transition is the second order in the figure.

\section{Dynamical Domain Wall by Black Hole}

It has been clear that the small black hole can create
the spherical EW domain wall by the local thermalization.
It is natural to ask whether
the spherical GUT domain wall could be formed by the similar process.
The similar analysis tells us that
it is impossible to create the stationary GUT wall
thermalized by the GUT interactions
because
the cross sections between particles at the GUT scale are too small
to thermalize the wall
and the lifetime of the black hole
whose Hawking temperature is greater than the GUT energy
is too short to keep the stationary wall.
Therefore
the thermalized spherical GUT domain wall
due to the local heat up and due to the thermal phase transition
does not appear.
However,
we can consider the formation of the GUT spherical wall
by the other mechanism
which does not require the local thermalization.

We will consider the spherical domain wall around a black hole
without the thermalization and
will discuss the dynamical stability of the wall
(see Figure \ref{NonThermal.eps}).
The radius of the spherical domain wall is smaller than the mean free
path because the system is not thermalized,
then the Hawking-radiated particles directly reach the domain wall.
When the particles obtain their mass from the Higgs field,
the particle climbing over the wall
gives a part of his momentum to the wall and
the reflected particle gives twice his momentum to the wall.
Then the domain wall should feel the outward pressure,
namely, the Hawking pressure.
On the other hand, the domain wall feels the inward pressure
because of the energy from the symmetric region and
the energy from the wall tension.
When the radius of the wall is $r$,
the Hawking pressure is proportional to $1/r^2$ and
the pressure from the tension is proportional to $-1/r$.
Therefore the balance among these pressures at a certain radius
results in
the existence of the dynamical stationary domain wall.
The condition for existence of the dynamical wall is depending
on the several parameters, e.g., mass of the particles,
degree of the freedom for the heavy particles, tension of the wall
and so on.
Generally speaking,
we can consider that the dynamical domain wall exists
when the Hawking temperature is greater than the critical temperature
of the phase transition.
By this mechanism for the pressure balance,
we conclude that the black hole whose Hawking temperature is greater
than the GUT scale creates the spherical GUT wall around it.
The structure of the dynamical wall can be obtained by
solving the motion equation for the Higgs field
and the particles radiated from the black hole \cite{Nagatani:2002aa}.

\section{Electroweak Baryon Number Creation by Black Hole}

The electroweak baryogenesis mechanism has been proposed by
Cohen, Kaplan and Nelson (CKN mechanism),
which uses the EW domain wall from the assumption for
the first order phase transition of EW theory \cite{Cohen:1991iu}.
We can consider the similar baryogenesis mechanism
by using the spherical EW domain wall around a black hole
\cite{Nagatani:1998rt, Nagatani:1998gv, Nagatani:2001nz}.
We summarize distinctions between the CKN mechanism and
our baryogenesis mechanism
as the Sakharov's three criteria for baryon number creation in Table 1.
  \begin{center}
   \begin{tabular}{@{\hspace{5mm}}l@{\hspace{5mm}}|c|c}
    \hline\hline
    Sakharov's criteria &
    \hspace{10mm} CKN mechanism \hspace{10mm} &
    \hspace{5mm} our mechanism \hspace{5mm} \\
    \hline\hline
    1) baryon \# violation & \multicolumn{2}{c}{sphaleron process} \\
    \hline
    2) C- and CP- violation & \multicolumn{2}{c}{SM and extension of Higgs sector} \\
    \hline
    3) out of equilibrium & 1st order transition & Hawking radiation\\
    \hline\hline
   \end{tabular}\\[1mm]
   (Table 1) Distinctions in the Sakharov's criteria.
  \end{center}
The essence of the difference between the mechanisms is how to realize
the out of the thermal equilibrium;
In the CKN mechanism,
the EW domain wall by the first order phase transition
at the cooling process of the universe is running through the
stationary plasma,
on the other hand,
the outgoing plasma-flow due to the Hawking radiation
goes across the stationary spherical wall in our model.
Our spherical EW domain wall
is classified into the {\it thick wall} in the CKN mechanism
because the wall is thermalized by the SM interactions,
therefore, the amount of baryon number production can be evaluated by
considering the {\it spontaneous baryogenesis mechanism}.
Finally, we have obtained the total baryon number created by a black hole
in his lifetime:
\begin{eqnarray*}
 B &\simeq& 10^{-9} \times \Delta\theta_\CP \; \frac{m_\BH}{T_\weak},
\end{eqnarray*}
where $m_\BH$ is the mass of the black hole,
$\theta_\CP$ is the CP-broken phase assumed in the wall
and $T_\weak \simeq 100\GeV$ is the critical temperature of the EW phase 
transition.

 \section{Cosmological Model with Primordial Black Holes}

\begin{figure}[htbp]%
\begin{center}%
 \includegraphics[scale=0.9]{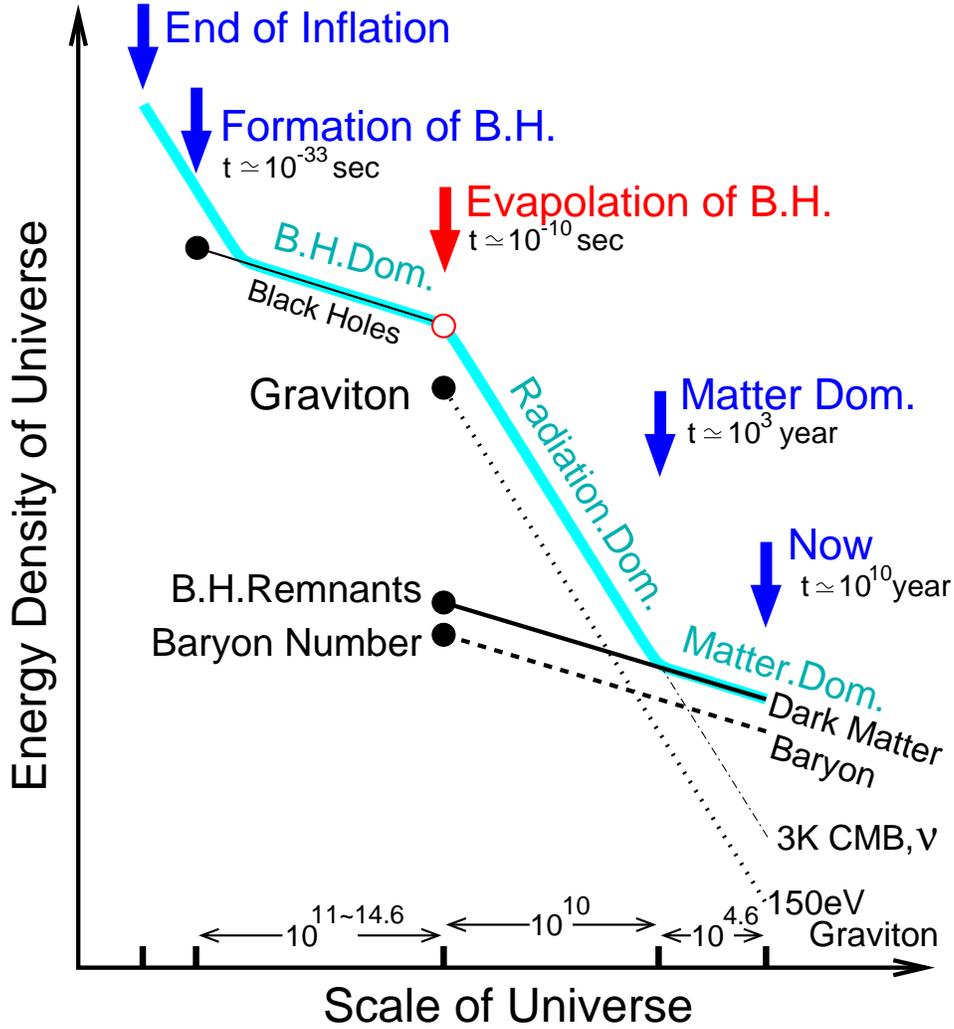}\hspace{5mm}%
 \caption{
 The density history of the universe in our model
 --- from the end of inflation to the present universe.
 Both axes are log scale.
 The thick curve means the critical density of the universe.
 1) The reheating at the end of the inflation
 makes the hot radiation dominant universe with the critical density.
 2) A characteristic density fluctuation created at the inflation
 forms the primordial black holes with mass of several hundred kilogram
 when the age of the universe is $10^{-33}\second$.
 The energy density of the primordial black holes should be greater than 
 $10^{-10}$ of the critical density.
 3) The universe becomes black hole dominant.
 4) When the age of the universe is $10^{-10}\second$ ,
 the primordial black holes evaporate
 with heating up the universe.
 The black holes create baryon number in the universe
 by our mechanism of the spherical EW domain wall,
 and also create gravitons.
 The Planck-remnants left after the evaporation of black holes
 behave as the cold dark matter.
 5) The history of the universe
 after the evaporation of the black holes in our model
 is the same as the ordinary models of cosmology.
 }%
 \label{DHistory.eps}%
\end{center}%
\end{figure}

We have discussed the phenomena caused by a black hole,
next, we will consider the applications of the phenomena to cosmology,
e.g., the baryogenesis, the cold dark matter and so on \cite{Nagatani:2001nz}.
We have constructed a cosmological model which can explain
the baryon-entropy ratio in the universe $B/S \sim 10^{-10}$ required
from the observations and the big-bang nucleosynthesis (BBN) theory.
Our model requires the following assumptions; 
(i) the universe earlier than $10^{-10}\second$ is dominated by the
primordial black holes with mass of the several hundred kilogram:
\begin{eqnarray*}
 m_\BH
  &\simeq&
  \frac{\xi}{20} \left(\frac{m_\planck}{T_\weak}\right)^{2/3} m_\planck
  \qquad (\xi = 1.0 \sim 5.2)
\end{eqnarray*}
and 
(ii) the spherical EW domain wall around any black hole has CP-broken phase
with order of one: $\Delta\theta_\CP \sim 1$.
In our model, the following scenario is supposed;
(1) the density-fluctuation with a spectrum peaked sharp is created
at the inflation era,
(2) the fluctuation creates the primordial black hole
with required mass after the end of inflation and
(3) when the age of the universe is $10^{-10}\second$,
the primordial black holes evaporate with creating baryon number
in the universe
and the universe is reheated up
by the Hawking radiation of the black holes (see Figure \ref{DHistory.eps}).
The primordial black holes created by the fluctuation
behave as the dust-like matter,
then the energy density of the black holes evolves
with the $(-3)$ power of the scale factor of the universe expansion.
On the other hand, the energy density of the radiation
created by the reheating at the end of the inflation
evolves with the $(-4)$ power of the scale factor.
Therefore
the expanding universe can become primordial black hole dominant
if the initial energy density of the created primordial black hole
is greater than $10^{-10}$ of the energy density of the universe.

\section{Black Hole Remnants as Cold Dark Matter}

Many authors have discussed
whether there exists a {\it remnant}
which is left after the evaporation of the black hole
by the Hawking radiation
and what is it if it exists
from the various viewpoints, e.g.,
the uncertainty principle, the information problems,
the corrections from quantum gravity, the BPS states and so on
\cite{Zeldovich, Bowick:1988xh, Coleman:1991sj, Gibbons:1987ps}.
The issue does not have been understood definitely
at the present time
because the quantum gravity does not have been completed.
Many authors discuss that, if it exists,
the mass of the remnant may be about the Planck mass,
namely, the {\it Planck remnant} is left.
Here, if any black holes in our model leave the Planck remnants,
how much energy density of them exists in the present universe?
The energy density of the Planck remnant is evaluated
as about ten times of that of the baryonic matter.
Typically the black hole with $300\kg$ creates
the baryonic matter with $3\microgram$ and
left the Planck remnant with $20\microgram$.
Our model predicts $\Omega_\remnant \sim 0.3$
which is consistent to the current observations.
Therefore
the remnant in our model can be a good candidate for the cold dark matter.

\section{Spontaneous Charging up of Black Hole}

At the final stage of the evaporation of the black hole,
the Hawking temperature increases rapidly
and it exceeds the critical temperature of the GUT phase transition.
Therefore the spherical GUT domain wall is formed dynamically
around such a black hole.
The dynamical domain wall is classified into the thin wall
in the CKN mechanism.

In the ordinary way
the Hawking radiation is essentially neutral for any kind of the
charges.
However
a black hole with a spherical thin wall radiates net (hyper-) charge
and the black hole is (hyper-) charging up itself
when the following assumptions are satisfied;
1) there is a CP-broken phase in the thin wall like the CKN model
and 2) the GUT contains heavy chiral (charge-assigned) fermions
like the top quarks in the SM \cite{Nagatani:2002aa}.
Because there is an asymmetry of the transparency/reflection rate
on the wall
between positive and negative (hyper) charge.
The mechanism,
{\it spontaneous charging up of the black hole},
is a variant application
of the {\it charge transport} mechanism in the CKN model,
so the (hyper) charge is transported to the black hole.

By the spontaneous charging up mechanism
we can expect that the remnant of the black hole has (hyper) charge
and it may be the extremal (hyper) charged black hole with a Planck mass.
The large-charged extremal black hole is changing to the smaller-charged
black hole with a smaller mass
by the Schwinger pair production and
the Hawking radiation associated with the charge loss.
We should wait the completion of the quantum gravity
to know the final of the transition sequence.

\section{Black Atom}

If the extremal electromagnetic charged black holes with a Planck mass
are the main contents of the cold dark matter,
how do they behave?
If the charge of the remnant $Z_\remnant$ is positive,
it will capture $Z_\remnant$ electrons and forms a kind of atom
when the temperature of the universe becomes
lower than about $0.3\eV$
like the recombination of the hydrogen-atom from the proton and the electron.
However the Planck-massive atom
with a positive electric charged remnant has been
excluded
as the main content of the cold dark matter by the several observations,
e.g., OYA, MICA and so on.

If the charge of the remnant is negative,
it captures $|Z_\remnant|$ protons and forms a structure like an atom,
we call it the {\it black atom}.
The nucleus of the black atom is the negative charged remnant
(which may be a extremal black hole)
and protons go around the nucleus
like the orbital electrons of the ordinary atom.
The ionization energy of the black atom is about 2000 times of
that of the ordinary atom
because mass of the proton is about 2000 times of that of the electron.
Therefore the black atoms are formed as 
the recombination of the remnants and the protons
in the universe with temperature about $0.6\keV$.
The Bohr radius of the black atom is
$1/2000$ of that of the ordinary atom,
the cross section is also smaller than the ordinary atom
and the present observations do not have excluded
existence of the black atom as the main content of the cold dark matter.

The extremal black hole as the nucleus of the black atom
can capture the orbital protons
and the extremal black hole becomes the smaller black hole.
The capturing cross section is about Planck area.
Therefore the black atom has finite lifetime and
the lifetime can be evaluated
by the similar way for the positronium-lifetime.
The black atom has a lifetime beyond the cosmological time $(10\Gyear)$
when $|Z_\remnant| \lnear 7$
with a point-particle-approximation for the proton.

\section{Cosmological Graviton Background}

The gravitons as well as the other particles
are radiated with the Hawking temperature from the black hole.
Therefore our model predicts the cosmological graviton background (CGB)
in the present universe,
which was created when the primordial black holes evaporated
\cite{Nagatani:2001nz}.
The energy density of the CGB is evaluated
as about $1/82$ of that of the cosmological microwave background (CMB).
The energy-spectrum of the CGB is much different from the
Planck-spectrum
and it has a peak at the energy $120 \sim 280 \eV$.

\section{Conclusions}

Our model requires the three assumptions;
(i) the black hole dominant,
(ii) the CP-broken phase in the wall with $O(1)$
and (iii) the Planck remnant of the black hole.
The assumption (i) requires the condition for the inflation models,
more concretely,
it requires that the reheating temperature at the end of the inflation
should be higher than about $10^{12}\GeV$ and so on.
The assumption (ii) gives us the restriction for
the extension of the SM.
B.~J.~Carr \etal has given us the restriction for the primordial black holes
in the early universe from the various theories and the observations
\cite{Carr:1994ar},
and the assumptions (i) is included in
the only one of the allowed parameter region of the black hole dominant.
If the remnants have only hyper charge,
the remnants are classified into the WIMP dark matter with Planck mass
and detection of them may be difficult.
If the remnants are the black atoms previously mentioned,
then the direct detection for them will be possible in the near future.

\Acknowledgements

We would like to thank K.~Kohri, T.~Sakai and K.~Shigetomi
for their useful discussions and suggestions.
We also appreciate helpful comments and advice of
A.~I.~Sanda, M.~Ninomiya and T. Nakamura.
YN is indebted to the Japan Society for
the Promotion of Science (JSPS) for its financial support.
The work is supported in part by a Grant-in-Aid for Scientific Research
from the Ministry of Education, Culture, Sports, Science and Technology
(No. 199903665).

\end{document}